\documentclass[aps,prl,showpacs,preprintnumbers,amsmath,amssymb,twocolumn]{revtex4}

\usepackage{graphicx}
\usepackage{dcolumn}
\usepackage{bm}

\begin{document}

\title{
Web-assisted tunneling in the kicked harmonic oscillator}

\author{Andr\'e R. R. Carvalho and Andreas Buchleitner}
\affiliation{Max-Planck-Institut f\"ur Physik komplexer Systeme,
N\"othnitzer Strasse 38, D-01187 Dresden}

\date{\today}

\begin{abstract}
We show that heating of harmonically trapped ions by periodic delta kicks 
is dramatically enhanced at isolated values of the Lamb-Dicke parameter. At
these values, quasienergy eigenstates localized on island structures undergo avoided
crossings with extended web-states. 
\end{abstract}


\maketitle

Controlling the state and the time evolution of quantum systems is one of the
central themes of current research in experimental and theoretical atomic
physics, quantum optics, and mesoscopics. Tailoring wave packets in Rydberg
systems \cite{weinacht98}, producing single photons on demand
\cite{brattke00}, creating coherent 
superpositions of macroscopic persistent-current states \cite{wal00}, and, most
recently, the 
controlled production of multiparticle entanglement \cite{roos04}, are
prominent examples of 
what is often coined as ``quantum state engineering'', such as to stress the almost
perfect control we have achieved on matter on the microscopic level. Whilst
many of these schemes still rely on an analysis of the quantum dynamics in
terms of some unperturbed basis of the quantum system under control, it has
become clear during the last decade that generic features of strongly coupled
(``complex'') quantum systems allow for novel and often extremely robust
strategies of quantum control. In such systems, studied in
much detail in the area of quantum chaos, peculiar eigenstates 
emerge which exhibit unexpected (and unexpectedly robust)  
localization properties and dynamics. Most prominent examples
thereof are nondispersive wave packets in periodically driven
quantum systems \cite{maeda04}, 
quantum resonances \cite{darcy01}, and quantum accelerator modes
\cite{oberthaler99}. A 
considerable part of 
these ``strong coupling'' quantum control schemes relies on some underlying
classical dynamics, which in general is mixed regular-chaotic, and 
precisely the rich structure of a mixed classical phase space is at their very
origin \cite{abu02,fishman03}. A large class of these systems follows the
Kolmogorov-Arnold-Moser 
(KAM) scenario \cite{lichtenberg83}, i.e. regular phase space structures  
are destroyed gradually as the coupling strength is increased. Yet, there is
another kind of classically chaotic dynamics, which goes under the name
``non-KAM'' chaos \cite{weakchaos}, where the phase space flow is 
fundamentally altered at arbitrarily weak perturbations. 
Recently, the consequences of such
non-KAM transitions have been observed in electron transport in superlattices
\cite{fromholdnat}, 
where enhanced transport accross the
lattice was observed, due to the sudden (non-KAM) appearence and disappearence
of unbounded 
stochastic web-structures in classical phase space, at well-defined values of some
control parameter. Hence, enhanced transport was enforced by
(controlled) abrupt
changes in the underlying classical phase space structure. 

Here we expose a
different pathway for the controlled enhancement of transport in another non-KAM
system, the kicked harmonic oscillator
\cite{nonlinear4,khoQC,rebuzzini,arrckho}, at {\em fixed} 
phase space structure, by simply tuning the effective value of $\hbar$. We
will see that, for a suitable choice of the initial oscillator state, dramatic
enhancement of the energy absorption by the oscillator from the kicking field
can be achieved due to avoided
crossings of localized (regular) with extended (web-like) eigenstates of the
kicked systems. We argue that
our scenario can be easily observed in state-of-the-art experiments on
harmonically trapped, kicked cold ions or atoms.   

The classical Hamiltonian of the kicked harmonic
 oscillator~\cite{weakchaos,chernikov87} describes a harmonically trapped (in
 1D, with trap frequency $\nu$) particle of mass $m$, subject to a one
 dimensional, spatially periodic potential (wave vector $k=2\pi/\lambda$,
 modulation depth $A$) which is periodically switched on and off at integer
 multiples of the kicking period $\tau$:   
\begin{equation}	
\label{classham}
H =  \frac{p^2}{2m} + \frac{m\nu^2 x^2}{2} + A \, \cos(k x) \sum_n
\delta(t-n\tau)\, .
\end{equation}
Further inspection of the Hamiltonian reveals that the classical
phase space structure is
completely determined by the parameters 
\begin{equation}
\label{parameters}
K= \frac{A k^2}{m \nu} \qquad \textrm {and} \qquad \alpha=\nu\tau \equiv
\frac{2\pi}{q}\, ,
\end{equation}
which define the stochasticity parameter and the ratio between kicking 
and oscillator period, respectively. This is a consequence of
the form-invariance of the equations of motion derived from (\ref{classham}),
under transformation to the scaled position and momentum coordinates $v=k x$
and $u=k p/m$.
 
A specific property of the kicked harmonic oscillator is that its phase space
is unbounded, thus allowing for transport to infinity (which, in the trapped
ion scenario, is tantamount to unbounded heating). Furthermore, it exhibits
peculiar symmetry properties determined by the ratio $q=(2\pi/\nu )/\tau$ of
oscillator and kicking period: For 
integer $q$, in addition to confined regular islands, it
displays 
a stochastic web (reaching out to infinity in phase space) 
with crystal ($q \in q_c\equiv\{3,\,4,\,6\}$) or
quasi-crystal symmetry ($q \notin \,q_c$), along which the system diffuses for
suitable initial conditions. 
This web is characterized not only by its symmetry but also by its thickness
that broadens (shrinks) as the value of 
the stochasticity parameter
$K$ increases (decreases). 

With the harmonic oscillator annihilation operator $\hat a = (\hat v+\hat
u)/2\eta$ and its hermitean conjugate $\hat a^{\dag}$ derived from the scaled
center of mass coordinates of the trapped particle, the quantum  version of~(\ref{classham}) reads
 \begin{equation}
\label{quantham}
\hat H = \hbar \nu \hat a^{\dag}\hat a + \hbar \tilde K
  \left\{\cos\left[\eta(\hat a + \hat a^{\dag})\right]
  \right\}\sum_{n=0}^\infty \delta(t-n\tau),
\end{equation}
with  
$\tilde K= K/2 \eta^2$, and the Lamb-Dicke parameter $\eta=k
\sqrt{\hbar/2 m \nu}$. Note that the latter measures the ratio of the width of the
harmonic oscillator ground state in units of the wave length of the kicking
potential, and that its squared value plays the role of an effective
Planck constant -- which can be tuned easily in ion trap experiments. 

Since (\ref{quantham}) is periodic in time, the time evolution of any initial
state $|\psi_0\rangle $ is given by the action of the one-cycle Floquet
propagator
\begin{equation}
\hat U \vert \varepsilon_j \rangle= e^{-i \alpha \hat a^{\dag} \hat a} e^{-i
\tilde K \cos\left[\eta(\hat a + \hat a^{\dag})\right] } \vert \varepsilon_j
\rangle = e^{i \phi_j}\vert \varepsilon_j \rangle,
\label{floq}
\end{equation}
with $\phi_j$ the quasi-energies and $\vert \varepsilon_j
\rangle$ the associated Floquet eigenstates. Due to 
the $\delta$-term in (\ref{quantham}), the Floquet propagator factorizes into
a free evolution and a kicking part, and inspection of the former shows that 
the above classical condition
for 
the emergence of a web structure in classical phase space is also quantum
mechanically distinguished, since integer $q$
is equivalent to a (fractional) revival condition \cite{averbukh89} 
for the free 
evolution right upon the subsequent kick. One therefore
expects quantum
signatures of web-sustained transport in classical phase space -- which we
shall 
now establish by combining dynamical and
spectral information. 

\begin{figure}
\includegraphics[width=7.0cm]{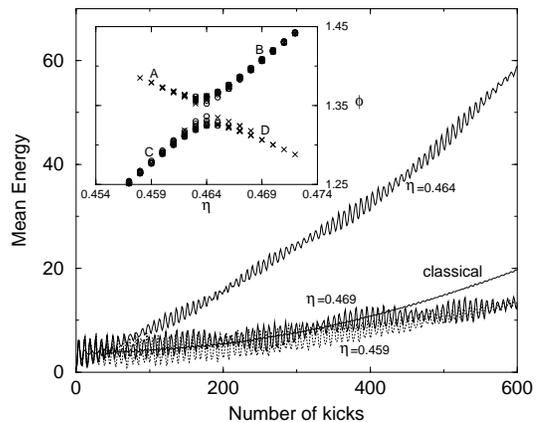}
\caption{Mean energy as a function of the number of kicks, for $q=6$, $K=2.0$, 
$\vert \psi_0 \rangle =\vert 0\rangle$, and Lamb-Dicke parameter $\eta=0.459$,
$0.464$ and $0.469$, at fixed classical phase space structure
(eq.~(\ref{parameters})). Also the classical time evolution is shown, for
comparison. The inset zooms into a given region of the level dynamics in the
lower panel of Fig.~\ref{levdyn} (denoted by an arrow there). Circles and crosses indicate eigenstates with
an overlap with the initial state $|\psi_0\rangle$ larger or smaller than
$10^{-2}$, respectively. Dramatic enhancement of the energy absorption by the
trapped particle from the kicking field is observed exactly at the center ($\eta=0.464)$ of the avoided crossing.} 
\label{dif}
\end{figure}

Let us start with Fig.~\ref{dif}, where we plot the mean energy 
of the kicked (quantum) particle, initially prepared in the harmonic
oscillator ground state, as a function of the number of kicks imparted
on it by the external field, for slightly different values of the Lamb-Dicke
parameter $\eta$, and for crystal symmetry $q=6$.
We see that a change of $\eta$ by merely one percent dramatically affects the
energy absorption by the particle from the field, eventually leading to
quantum transport which is much more efficient than the classical excitation
process. Note that the latter is unaffected by such tiny changes of $\eta$, 
since the results
displayed in the figure are obtained for fixed $K$ and $q$, hence for fixed
phase space structure! 

\begin{figure}
\includegraphics[width=8.0cm]{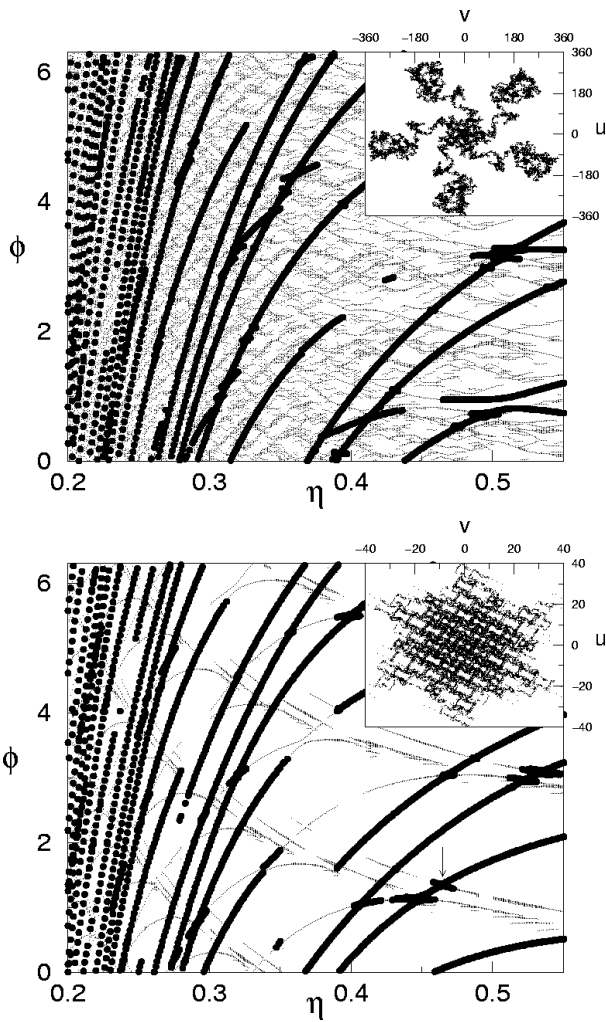}
\caption{Spectrum of the Floquet operator (\ref{floq}) as a function of the
Lamb-Dicke parameter $\eta$,
for $K=2.0$, $q=5$ (top) and $q=6$ (bottom). Only 
eigenstates with an overlap larger than
$10^{-3}$ with the initial state $|\psi_0\rangle$ are represented. Filled
circles represent $|\psi_0\rangle =|0\rangle$, while dots refer to a displaced vacuum state centered at $(1.3,3.0)$ (top) and $(1.2,2.0)$ (bottom). The insets
show the classical phase space explored by a single trajectory after 40000 kicks, when launched at a point near the
origin, for the same $K$ and $q$. The arrow in the bottom plot refers to the inset of Fig.~\ref{dif}.}
\label{levdyn}
\end{figure}

Whilst this latter observation appears puzzling
on a first glance, it is readily resolved by inspection of the evolution of
the quasienergy spectrum of (\ref{floq}) as a function of $\eta$. A global
view thereof is provided by Fig.~\ref{levdyn}, for quasicrystal ($q=5$) and
crystal ($q=6$)
symmetries, respectively, and otherwise the same parameters as in
Fig.~\ref{dif}.  
Only those eigenvalues associated with
eigenstates which have an overlap larger than $10^{-3}$ with $|\psi_0\rangle$
are shown. For both symmetries, two distinct initial states
$|\psi_0\rangle$ are considered: $|\psi_0\rangle =|0\rangle$ (circles), and 
$|0\rangle$ displaced to the points $(1.2,2.0)$ and
 $(1.3,3.0)$, located in the chaotic region of phase space, for $q=6$ and
$q=5$ (dots), respectively.  

We see that, for the vacuum initial state, the level dynamics
are rather similar in the
quasicrystal and in the crystal case, what is just the quantum signature of
the locally similar phase space structures in the vicinity of the origin. In
contrast, they are quite distinct for displaced
initial conditions, since, under these premises, the quantum dynamics probes
the rather distinct web structures of the underlying classical phase space
(illustrated by the insets in the figure, which show how a single trajectory
explores phase space in the long time limit -- note the different scales!). 
Independently of symmetry and initial condition, an abundance of avoided
crossings is apparent from both plots -- a typical signature of
the strong coupling between the driving field and the center of mass motion of
the trapped particle. Also note that, in the crystal case with
$|\psi_0\rangle $ in the chaotic domain, some avoided crossings 
are 
organized in a regular manner, what strongly suggests a
semiclassical 
origin -- which, however, is not relevant for our present purpose.

For the vacuum initial state, both
symmetries exhibit isolated or overlapping avoided crossings (of variable
size) at well defined
values of $\eta$, which turn out to cause the enhanced quantum transport
observed in Fig.~\ref{dif}: The inset in Fig.~\ref{dif} zooms into the level
structure around the avoided crossing at 
$(\eta=0.464;\phi =1.35)$ indicated by an arrow in the lower panel of Fig.~\ref{levdyn}.
Here, two quasienergies (as a matter of fact, two quasienergy bands
\cite{ABC}, due to 
the high spectral degeneracy induced by the symmetries of the Hamiltonian)
undergo an isolated avoided crossing, with maximal tunneling coupling between
the associated eigenstates at its center at $\eta=0.464$. Note from
Fig.~\ref{dif} that enhanced quantum transport is observed {\em precisely} at
this value. 
In contrast, for only
slightly smaller or larger values of the Lamb-Dicke parameter, i.e. slightly shifted with respect to the center of the avoided
crossing, the trapped particle absorbs energy with a much smaller rate, rather
similar to the classical energy absorption. Hence, the width of isolated avoided
crossings induces the sharp response of the system to tiny changes in $\eta$,
opening a narrow parameter window in which the system can efficiently tunnel
from its initial state $|\psi_0\rangle$ into another eigenstate which mediates
rapid transport.

\begin{figure}
\includegraphics[width=8.6cm]{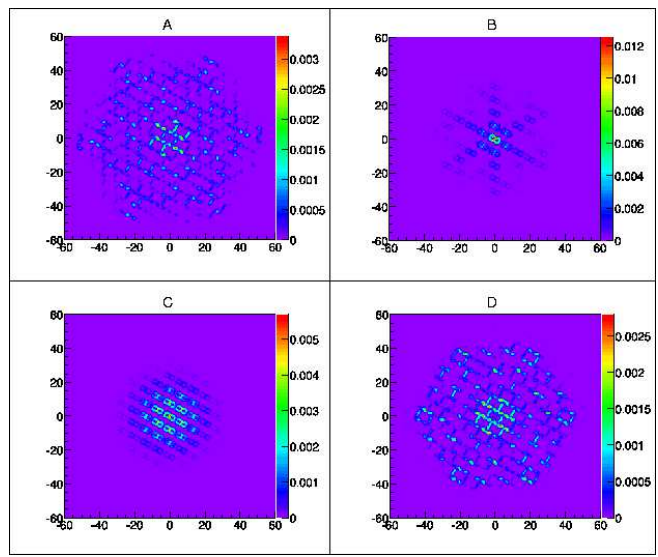}
\caption{(Color online) Husimi representations for the eigenstates associated with the points
A, B, C and D of the avoided crossing in the inset of Fig.~\ref{dif}. Phase
space is spanned by the coordinates $v/2\eta$ and $u/2\eta$, respectivley.
Clearly, the eigenstates localized in the vicinity of the hyperbolic point at
the origin (B and C) dominate the dynamics in the asymptotic
part ($\eta=0.459$ and $\eta=0.469$) of the anticrossing in the inset of
Fig.~\ref{dif}, whilst extended states localized on the stochastic web (A and D)
exhibit equal weight  at the center of the crossing 
parameter window for enhanced quantum transport.  
}
\label{eigenfunc}
\end{figure}

But which eigenstates of the kicked harmonic oscillator promote enhanced
diffusion? Let us inspect the phase space projection of the anticrossing
states in the inset of Fig.~\ref{dif}, ``on the left'' (labels A and C) and
``on the right'' (labels B and D) of the crossing's center, which are represented by their
Husimi functions in Fig.~\ref{eigenfunc}. 
As we approach the crossing from small values of $\eta$, the lower branch of
the anticrossing in the inset of Fig.~\ref{dif} has the largest overlap with
$|\psi_0\rangle$. The corresponding
eigenstate for $\eta=0.459$ is plotted 
in the lower left panel of Fig.~\ref{eigenfunc}, and is clearly localized on
the hyperbolic fixed point at the origin, and on its first replica on the
crystal lattice. It anticrosses with the state (A) shown in the upper left panel of the figure. Clearly, this latter state is extended (note that it has large
densities far away from the origin), localized on the
stochastic web, reaching out to the effective boundary of the phase space
which is defined by the finite size of the basis which we use for the
diagonalization of $\hat U$. Since such web states -- by their very nature --
can never be converged numerically, we carefully checked that the associated
eigenvalues saturate at a finite basis size, and that their 
localization properties within a finite phase space domain remain structurally
invariant when increasing the numerical basis. On the other side of the
avoided crossing, at $\eta=0.469$, the anticrossing states have exchanged their
localization properties, as illustrated in the right panels of
Fig.~\ref{eigenfunc}, and as to be expected from the usual anticrossing
scenario: Now the upper branch of the crossing provides the strongest support
for $|\psi_0\rangle$, and the associated eigenstate (B) is once 
again localized on the hyperbolic fixed point at the origin, whereas the lower
branches' eigenstate (D) is localized on the web. 

\begin{figure}
\includegraphics[width=6.0cm,angle=-90]{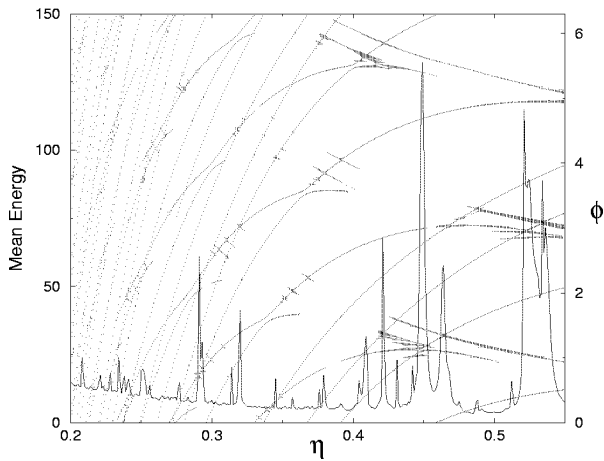}
\caption{Mean energy (left y-axis) 
after 600 kicks, as a function of the
Lamb-Dicke parameter $\eta$, and for fixed classical phase space structure,
with $K=2.0$ and $q=6$. The signal is strongly enhanced at those
values of $\eta$ which define avoided crossings between localized and web-like
eigenstates. 
} 
\label{fluct}
\end{figure}

Consequently, the web eigenstates
of the kicked harmonic oscillator lend support for enhanced quantum transport
as observed in Fig.~\ref{dif}, at sharply defined values of the effective
Planck constant (parametrized by the Lamb-Dicke parameter), where
near-resonant tunneling from the hyperbolic fixed point onto the web becomes
possible. Since the Lamb-Dicke parameter can be tuned by variation of the trap
frequency $\nu$ or of the kicking lattice's wavelength $2\pi/k$, such
web-enhanced tunneling from the harmonic oscillator ground state will have a
marked signature on the experimentally observed heating rate vs. $\eta$, as
illustrated in Fig.~\ref{fluct}.
Note that each of the peaks in the mean energy signal (after a fixed number of
kicks) plotted vs. $\eta$ in
this plot corresponds to anticrossings (or, for the broader resonance
structures, to overlapping avoided crossings) between localized and extended
states alike the one identified in
Figs.~\ref{dif} and \ref{eigenfunc}, at fixed phase space structure. 

Tuning
the Lamb Dicke parameter (at easily variable interaction time!) 
thus allows for sensitive probing of the local
density of states by an experimentally robust observable, and, vice versa, for
{\em efficient control} of the diffusion properties of the complex quantum system
under study. Also note that Fig.~\ref{fluct} is reminiscent of conductance
fluctuations which occur in solids \cite{kramer93}, mesoscopic systems
\cite{pichard90}, and 
strongly driven 
atoms \cite{krug03} -- which bear experimentally accessible information on the
spectral 
structure that supports the detected probability current. 

\bibliography{literatur04}

\begin{thebibliography}{22}
\expandafter\ifx\csname natexlab\endcsname\relax\def\natexlab#1{#1}\fi
\expandafter\ifx\csname bibnamefont\endcsname\relax
  \def\bibnamefont#1{#1}\fi
\expandafter\ifx\csname bibfnamefont\endcsname\relax
  \def\bibfnamefont#1{#1}\fi
\expandafter\ifx\csname citenamefont\endcsname\relax
  \def\citenamefont#1{#1}\fi
\expandafter\ifx\csname url\endcsname\relax
  \def\url#1{\texttt{#1}}\fi
\expandafter\ifx\csname urlprefix\endcsname\relax\def\urlprefix{URL }\fi
\providecommand{\bibinfo}[2]{#2}
\providecommand{\eprint}[2][]{\url{#2}}

\bibitem[{\citenamefont{Weinacht et~al.}(1998)\citenamefont{Weinacht, Ahn, and
  Bucksbaum}}]{weinacht98}
\bibinfo{author}{\bibfnamefont{T.~C.} \bibnamefont{Weinacht}},
  \bibinfo{author}{\bibfnamefont{J.}~\bibnamefont{Ahn}}, \bibnamefont{and}
  \bibinfo{author}{\bibfnamefont{P.~H.} \bibnamefont{Bucksbaum}},
  \bibinfo{journal}{Phys. Rev. Lett.} \textbf{\bibinfo{volume}{80}},
  \bibinfo{pages}{5508} (\bibinfo{year}{1998}).

\bibitem[{\citenamefont{Brattke et~al.}(2000)\citenamefont{Brattke, Varcoe, and
  Walther}}]{brattke00}
\bibinfo{author}{\bibfnamefont{S.}~\bibnamefont{Brattke}},
  \bibinfo{author}{\bibfnamefont{B.~T.~H.} \bibnamefont{Varcoe}},
  \bibnamefont{and} \bibinfo{author}{\bibfnamefont{H.}~\bibnamefont{Walther}},
  \bibinfo{journal}{Phys. Rev. Lett.} \textbf{\bibinfo{volume}{86}},
  \bibinfo{pages}{3534} (\bibinfo{year}{2000}).

\bibitem[{\citenamefont{van~der Wal et~al.}(2000)\citenamefont{van~der Wal, ter
  Haar, Wilhelm, Schouten, Harmans, Orlando, Lloyd, and Mooij}}]{wal00}
\bibinfo{author}{\bibfnamefont{C.}~\bibnamefont{van~der Wal}},
  \bibinfo{author}{\bibfnamefont{A.}~\bibnamefont{ter Haar}},
  \bibinfo{author}{\bibfnamefont{F.}~\bibnamefont{Wilhelm}},
  \bibinfo{author}{\bibfnamefont{R.}~\bibnamefont{Schouten}},
  \bibinfo{author}{\bibfnamefont{C.}~\bibnamefont{Harmans}},
  \bibinfo{author}{\bibfnamefont{T.}~\bibnamefont{Orlando}},
  \bibinfo{author}{\bibfnamefont{S.}~\bibnamefont{Lloyd}}, \bibnamefont{and}
  \bibinfo{author}{\bibfnamefont{J.}~\bibnamefont{Mooij}},
  \bibinfo{journal}{Science} \textbf{\bibinfo{volume}{290}},
  \bibinfo{pages}{773} (\bibinfo{year}{2000}).

\bibitem[{\citenamefont{Roos et~al.}(2004)\citenamefont{Roos, Riebe, H\"affner,
  H\"ansel, Benhelm, Lancaster, Becher, Schmidt-Kaler, and Blatt}}]{roos04}
\bibinfo{author}{\bibfnamefont{C.}~\bibnamefont{Roos}},
  \bibinfo{author}{\bibfnamefont{M.}~\bibnamefont{Riebe}},
  \bibinfo{author}{\bibfnamefont{H.}~\bibnamefont{H\"affner}},
  \bibinfo{author}{\bibfnamefont{W.}~\bibnamefont{H\"ansel}},
  \bibinfo{author}{\bibfnamefont{J.}~\bibnamefont{Benhelm}},
  \bibinfo{author}{\bibfnamefont{G.}~\bibnamefont{Lancaster}},
  \bibinfo{author}{\bibfnamefont{C.}~\bibnamefont{Becher}},
  \bibinfo{author}{\bibfnamefont{F.}~\bibnamefont{Schmidt-Kaler}},
  \bibnamefont{and} \bibinfo{author}{\bibfnamefont{R.}~\bibnamefont{Blatt}},
  \bibinfo{journal}{Science} \textbf{\bibinfo{volume}{304}},
  \bibinfo{pages}{1478} (\bibinfo{year}{2004}).

\bibitem[{\citenamefont{Maeda and Gallagher}(2004)}]{maeda04}
\bibinfo{author}{\bibfnamefont{H.}~\bibnamefont{Maeda}} \bibnamefont{and}
  \bibinfo{author}{\bibfnamefont{T.~F.} \bibnamefont{Gallagher}},
  \bibinfo{journal}{Phys. Rev. Lett.} \textbf{\bibinfo{volume}{92}},
  \bibinfo{pages}{133004} (\bibinfo{year}{2004}).

\bibitem[{\citenamefont{d'Arcy et~al.}(2001)\citenamefont{d'Arcy, Godun,
  Oberthaler, Cassettari, and Summy}}]{darcy01}
\bibinfo{author}{\bibfnamefont{M.~B.} \bibnamefont{d'Arcy}},
  \bibinfo{author}{\bibfnamefont{R.~M.} \bibnamefont{Godun}},
  \bibinfo{author}{\bibfnamefont{M.~K.} \bibnamefont{Oberthaler}},
  \bibinfo{author}{\bibfnamefont{D.}~\bibnamefont{Cassettari}},
  \bibnamefont{and} \bibinfo{author}{\bibfnamefont{G.~S.} \bibnamefont{Summy}},
  \bibinfo{journal}{Phys. Rev. Lett.} \textbf{\bibinfo{volume}{87}},
  \bibinfo{pages}{074102} (\bibinfo{year}{2001}).

\bibitem[{\citenamefont{Oberthaler et~al.}(1999)\citenamefont{Oberthaler,
  Godun, d'Arcy, Summy, and Burnett}}]{oberthaler99}
\bibinfo{author}{\bibfnamefont{M.~K.} \bibnamefont{Oberthaler}},
  \bibinfo{author}{\bibfnamefont{R.~M.} \bibnamefont{Godun}},
  \bibinfo{author}{\bibfnamefont{M.~B.} \bibnamefont{d'Arcy}},
  \bibinfo{author}{\bibfnamefont{G.~S.} \bibnamefont{Summy}}, \bibnamefont{and}
  \bibinfo{author}{\bibfnamefont{K.}~\bibnamefont{Burnett}},
  \bibinfo{journal}{Phys. Rev. Lett.} \textbf{\bibinfo{volume}{83}},
  \bibinfo{pages}{4447} (\bibinfo{year}{1999}).

\bibitem[{\citenamefont{Buchleitner et~al.}(2002)\citenamefont{Buchleitner,
  Delande, and Zakrzewski}}]{abu02}
\bibinfo{author}{\bibfnamefont{A.}~\bibnamefont{Buchleitner}},
  \bibinfo{author}{\bibfnamefont{D.}~\bibnamefont{Delande}}, \bibnamefont{and}
  \bibinfo{author}{\bibfnamefont{J.}~\bibnamefont{Zakrzewski}},
  \bibinfo{journal}{Phys. Rep.} \textbf{\bibinfo{volume}{368}},
  \bibinfo{pages}{409} (\bibinfo{year}{2002}).

\bibitem[{\citenamefont{Fishman et~al.}(2003)\citenamefont{Fishman, Guarneri,
  and Rebuzzini}}]{fishman03}
\bibinfo{author}{\bibfnamefont{S.}~\bibnamefont{Fishman}},
  \bibinfo{author}{\bibfnamefont{I.}~\bibnamefont{Guarneri}}, \bibnamefont{and}
  \bibinfo{author}{\bibfnamefont{L.}~\bibnamefont{Rebuzzini}},
  \bibinfo{journal}{J. Stat. Phys.} \textbf{\bibinfo{volume}{110}},
  \bibinfo{pages}{911} (\bibinfo{year}{2003}).

\bibitem[{\citenamefont{Lichtenberg and Lieberman}(1983)}]{lichtenberg83}
\bibinfo{author}{\bibfnamefont{A.~J.} \bibnamefont{Lichtenberg}}
  \bibnamefont{and} \bibinfo{author}{\bibfnamefont{M.~A.}
  \bibnamefont{Lieberman}}, \emph{\bibinfo{title}{Regular and Stochastic
  Motion}}, vol.~\bibinfo{volume}{38} of \emph{\bibinfo{series}{Applied
  Mathematical Sciences}} (\bibinfo{publisher}{Springer},
  \bibinfo{address}{Berlin}, \bibinfo{year}{1983}).

\bibitem[{\citenamefont{Zaslavsky et~al.}(1992)\citenamefont{Zaslavsky,
  Sagdeev, Usikov, and Chernikov}}]{weakchaos}
\bibinfo{author}{\bibfnamefont{G.}~\bibnamefont{Zaslavsky}},
  \bibinfo{author}{\bibfnamefont{R.}~\bibnamefont{Sagdeev}},
  \bibinfo{author}{\bibfnamefont{D.}~\bibnamefont{Usikov}}, \bibnamefont{and}
  \bibinfo{author}{\bibfnamefont{A.}~\bibnamefont{Chernikov}},
  \emph{\bibinfo{title}{Weak chaos and quasi-regular patterns}}
  (\bibinfo{publisher}{Cambridge University Press}, \bibinfo{year}{1992}).

\bibitem[{\citenamefont{Fromhold et~al.}(2004)\citenamefont{Fromhold,
  Patan{\`e}, Bujkiewicz, Wilkinson, Fowler, Sherwood, Stapleton, Krokhin,
  eaves, Henini et~al.}}]{fromholdnat}
\bibinfo{author}{\bibfnamefont{T.}~\bibnamefont{Fromhold}},
  \bibinfo{author}{\bibfnamefont{A.}~\bibnamefont{Patan{\`e}}},
  \bibinfo{author}{\bibfnamefont{S.}~\bibnamefont{Bujkiewicz}},
  \bibinfo{author}{\bibfnamefont{P.~B.} \bibnamefont{Wilkinson}},
  \bibinfo{author}{\bibfnamefont{D.}~\bibnamefont{Fowler}},
  \bibinfo{author}{\bibfnamefont{D.}~\bibnamefont{Sherwood}},
  \bibinfo{author}{\bibfnamefont{S.~P.} \bibnamefont{Stapleton}},
  \bibinfo{author}{\bibfnamefont{A.~A.} \bibnamefont{Krokhin}},
  \bibinfo{author}{\bibfnamefont{L.}~\bibnamefont{eaves}},
  \bibinfo{author}{\bibfnamefont{M.}~\bibnamefont{Henini}},
  \bibnamefont{et~al.}, \bibinfo{journal}{Nature}
  \textbf{\bibinfo{volume}{428}}, \bibinfo{pages}{726} (\bibinfo{year}{2004}).

\bibitem[{\citenamefont{Berman et~al.}(1991)\citenamefont{Berman, Rubaev, and
  Zaslavsky}}]{nonlinear4}
\bibinfo{author}{\bibfnamefont{G.}~\bibnamefont{Berman}},
  \bibinfo{author}{\bibfnamefont{V.}~\bibnamefont{Rubaev}}, \bibnamefont{and}
  \bibinfo{author}{\bibfnamefont{G.}~\bibnamefont{Zaslavsky}},
  \bibinfo{journal}{Nonlinearity} \textbf{\bibinfo{volume}{4}},
  \bibinfo{pages}{543} (\bibinfo{year}{1991}).

\bibitem[{\citenamefont{Shepelyansky and Sire}(1992)}]{khoQC}
\bibinfo{author}{\bibfnamefont{D.}~\bibnamefont{Shepelyansky}}
  \bibnamefont{and} \bibinfo{author}{\bibfnamefont{C.}~\bibnamefont{Sire}},
  \bibinfo{journal}{Europhys. Lett.} \textbf{\bibinfo{volume}{20}},
  \bibinfo{pages}{95} (\bibinfo{year}{1992}).

\bibitem[{\citenamefont{Borgonovi and Rebuzzini}(1995)}]{rebuzzini}
\bibinfo{author}{\bibfnamefont{F.}~\bibnamefont{Borgonovi}} \bibnamefont{and}
  \bibinfo{author}{\bibfnamefont{L.}~\bibnamefont{Rebuzzini}},
  \bibinfo{journal}{Phys. Rev. E} \textbf{\bibinfo{volume}{52}},
  \bibinfo{pages}{2302} (\bibinfo{year}{1995}).

\bibitem[{\citenamefont{Carvalho et~al.}(2004)\citenamefont{Carvalho,
  de~Matos~Filho, and Davidovich}}]{arrckho}
\bibinfo{author}{\bibfnamefont{A.~R.~R.} \bibnamefont{Carvalho}},
  \bibinfo{author}{\bibfnamefont{R.~L.} \bibnamefont{de~Matos~Filho}},
  \bibnamefont{and}
  \bibinfo{author}{\bibfnamefont{L.}~\bibnamefont{Davidovich}},
  \bibinfo{journal}{Phys. Rev. E} \textbf{\bibinfo{volume}{70}},
  \bibinfo{pages}{026211} (\bibinfo{year}{2004}).

\bibitem[{\citenamefont{Chernikov et~al.}(1987)\citenamefont{Chernikov,
  Sagdeev, Usikov, Zakharov, and Zaslavsky}}]{chernikov87}
\bibinfo{author}{\bibfnamefont{A.~A.} \bibnamefont{Chernikov}},
  \bibinfo{author}{\bibfnamefont{R.~Z.} \bibnamefont{Sagdeev}},
  \bibinfo{author}{\bibfnamefont{D.~A.} \bibnamefont{Usikov}},
  \bibinfo{author}{\bibfnamefont{M.~Y.} \bibnamefont{Zakharov}},
  \bibnamefont{and} \bibinfo{author}{\bibfnamefont{G.~M.}
  \bibnamefont{Zaslavsky}}, \bibinfo{journal}{Nature}
  \textbf{\bibinfo{volume}{326}}, \bibinfo{pages}{559} (\bibinfo{year}{1987}).

\bibitem[{\citenamefont{Averbukh and Perelman}(1989)}]{averbukh89}
\bibinfo{author}{\bibfnamefont{I.~S.} \bibnamefont{Averbukh}} \bibnamefont{and}
  \bibinfo{author}{\bibfnamefont{N.~F.} \bibnamefont{Perelman}},
  \bibinfo{journal}{Phys. Lett. A} \textbf{\bibinfo{volume}{139}},
  \bibinfo{pages}{449} (\bibinfo{year}{1989}).

\bibitem[{\citenamefont{Ketzmerick et~al.}(1998)\citenamefont{Ketzmerick,
  Kruse, and Geisel}}]{ABC}
\bibinfo{author}{\bibfnamefont{R.}~\bibnamefont{Ketzmerick}},
  \bibinfo{author}{\bibfnamefont{K.}~\bibnamefont{Kruse}}, \bibnamefont{and}
  \bibinfo{author}{\bibfnamefont{T.}~\bibnamefont{Geisel}},
  \bibinfo{journal}{Phys. Rev. Lett.} \textbf{\bibinfo{volume}{80}},
  \bibinfo{pages}{137} (\bibinfo{year}{1998}).

\bibitem[{\citenamefont{Kramer and MacKinnon}(1993)}]{kramer93}
\bibinfo{author}{\bibfnamefont{B.}~\bibnamefont{Kramer}} \bibnamefont{and}
  \bibinfo{author}{\bibfnamefont{A.}~\bibnamefont{MacKinnon}},
  \bibinfo{journal}{Rep. Prog. Phys.} \textbf{\bibinfo{volume}{56}},
  \bibinfo{pages}{1469} (\bibinfo{year}{1993}).

\bibitem[{\citenamefont{Pichard et~al.}(1990)\citenamefont{Pichard, Zanon,
  Imry, and Stone}}]{pichard90}
\bibinfo{author}{\bibfnamefont{J.-L.} \bibnamefont{Pichard}},
  \bibinfo{author}{\bibfnamefont{N.}~\bibnamefont{Zanon}},
  \bibinfo{author}{\bibfnamefont{Y.}~\bibnamefont{Imry}}, \bibnamefont{and}
  \bibinfo{author}{\bibfnamefont{A.~D.} \bibnamefont{Stone}},
  \bibinfo{journal}{J. Phys. (France)} \textbf{\bibinfo{volume}{51}},
  \bibinfo{pages}{587} (\bibinfo{year}{1990}).

\bibitem[{\citenamefont{Krug et~al.}(2003)\citenamefont{Krug, Wimberger, and
  Buchleitner}}]{krug03}
\bibinfo{author}{\bibfnamefont{A.}~\bibnamefont{Krug}},
  \bibinfo{author}{\bibfnamefont{S.}~\bibnamefont{Wimberger}},
  \bibnamefont{and}
  \bibinfo{author}{\bibfnamefont{A.}~\bibnamefont{Buchleitner}},
  \bibinfo{journal}{Eur. Phys. J. D} \textbf{\bibinfo{volume}{26}},
  \bibinfo{pages}{21} (\bibinfo{year}{2003}).

\end{thebibliography}

\end{document}